\documentclass[10pt]{article}
\usepackage{fullpage}
\usepackage{amsmath,amsfonts,amsthm,amssymb}
\usepackage{verbatim} 
\usepackage{hyperref}

\numberwithin{equation}{section}

\newcommand{\Hld}{\mathcal{H}}
\newcommand{\Dst}{\mathcal{D}}

\def\be{\begin{eqnarray}}
\def\ee{\end{eqnarray}}

\def\b*{\begin{eqnarray*}}
\def\e*{\end{eqnarray*}}

\makeatletter \@addtoreset{equation}{section}

\def\={\;=\;}
\def\.{\;.}

\def\1{\mathds{1}}

 \def\normeL2#1{\left\|{#1}\right\|_{L^2}}

\newcommand {\Chi} {{\bf \raise 1.5pt \hbox{$\delta$}}}
\newcommand{\footnotedva}[1]{}

\newtheorem*{lemma*}{Lemma}
\newtheorem*{prop*}{Proposition}
\newtheorem*{cor*}{Corollary}

\newtheorem*{remark*}{Remark}

\begin{document}

\title{Balancing small fixed and proportional transaction cost in trading strategies}
\author{Jose V. Alcala~and Arash Fahim\\
University of Michigan\\
alcala@umich.edu, fahimara@umich.edu}

\maketitle

\begin{abstract}
Transaction costs appear in financial markets in more than one form. There are several results in the literature on small proportional transaction cost and not that many on fixed transaction cost. In the present work, we heuristically study the effect of both types of transaction cost by focusing on a portfolio optimization. Here we assume the presence of fixed transaction cost and that there is a balance between fixed and proportional transaction cost, such that none of them dominates the other, asymptotically. We find out that the deviation of value function, when the fixed transaction cost is $\varepsilon$, from the Merton value function, without transaction cost, is of order $\varepsilon^\frac{1}{2}$ which is different from the pure proportional cost of $\varepsilon^\frac{2}{3}$. Based on this, we propose an expansion for the value function in terms of powers of $\varepsilon^\frac{1}{2}$. 
\end{abstract}


\section{Introduction.}

Trading strategies developed by Atkinson, Pliska, and Wilmott \cite{Atkinson} consider the effects of a small \emph{fixed cost} payed by the investor independently of the volume of the transaction. The case of small cost \emph{proportional} to the volume of the transaction has been studied by Shreve and Soner  \cite{shreve_soner} and Goodman and Ostrov \cite{Goodman}. In both cases the emphasis is on the asymptotic behavior of the trading strategies as the transaction cost vanishes and the problem approaches the \emph{idealized} no transaction cost problem studied by Merton \cite{merton}. In this paper we consider the case when the investor faces \emph{both} fixed and proportional vanishing transaction costs. We find the balance between the two costs that makes no cost dominant and describe the asymptotic optimal trading strategy as a function of the proportional cost parameter, $\lambda$, and the fixed cost parameter, $\varepsilon$.

The case of small proportional cost has been extensively studied. Shreve and Soner in \cite{shreve_soner} and Janecek and Shreve in \cite{janecek_shreve} present rigorous arguments based on viscosity solutions to show that in the case of power utility and a single stock, the value function of  the investment-consumption problem under small proportional cost $\lambda$ converges to the Merton value function as fast as $O(\lambda^\frac{2}{3})$. Rogers \cite{rogers}, proves the same result using a more probabilistic argument. For general utilities and multiple stocks, in \cite{atkinson_mokkhavesa}, the authors provide the result using a heuristic argument based on a perturbation method. Goodman and Ostrov \cite{Goodman} show the same result based on a dual probabilistic argument which also provides a clear understanding of the quasi-steady state probability density of the optimal portfolio. The traditional asymptotic expansion for the value function , $f$, is given by the equation
\begin{eqnarray}\label{expansionProportional}
f=f^0+\lambda^\frac{2}{3}f^2+\lambda f^3+\lambda^\frac{4}{3}f^4+o(\lambda^\frac{4}{3})\:.
\end{eqnarray}
In the recent paper \cite{soner_touzi}, the authors give a rigorous proof of the equation $|f-f^0|\sim O(\lambda^\frac{2}{3})$ for general utilities under some assumptions on the regularity of Merton value function and on the Merton ratio. Their argument are based on the theory of viscosity solutions.

There are fewer studies in the case of small fixed cost. Atkinson, Pliska and Wilmott  \cite{Atkinson} consider the effect on trading strategies of a small fixed cost  payed by the investor.  They perform a formal asymptotic analysis of the value function associated with the problem at hand and obtain an expansion of the form
\begin{equation*}
	\begin{split}
	f & = f_0 + \varepsilon f^{4} + o(\varepsilon)\:.
	\end{split}
\end{equation*}

We study the case of a single stock. The argument follows the heuristics used by Goodman and Ostrov \cite{Goodman} where the effect of the costs is decomposed in \emph{trading cost} and \emph{oportunity cost}, which is due to deviations from the idealized portfolio. The (approximate) optimal strategy keeps the portfolio position inside a \emph{hold region}, $\Hld$, centered around the idealized portfolio position. A transaction takes place only when the portfolio's position is on the boundary of $\Hld$ to move the portfolio position to the boundary of a \emph{destination region}, $\Dst$, also centered around the idealized portfolio position and contained in $\Hld$. Assuming that the destination region is fixed, an expansion of the hold region decreases the trading cost but increases the opportunity cost. On the other hand, if the hold region is fixed, an expansion of the destination region will certainly increase the opportunity cost but it is not \emph{a priori} clear how it affects the trading cost. The optimal trading strategy is obtained by quantifying and balancing these effects. The main tool the is equilibrium probability density for the portfolio position, $u$, which turns out to be constant in $\Dst$ and linear in $\Hld\setminus \Dst$.

The heuristic analysis has been shown to be the dual equivalent of the perturbation theory applied to $f$, the value function for the expected utility of the portfolio at a final time $T$, in the case of proportional transaction cost. In section \ref{FixedCost} we perform both the heuristic analysis and the perturbation analysis of the value function $f$ in the fixed transaction cost setting as a sanity check. In section \ref{Both} we rely only on the dual heuristics when both fixed and proportional costs are present. When the hold (destination) region has a radius equal to $\gamma$ ($\eta$) the opportunity cost is proportional to $\gamma^2+\eta^2$ and the transaction cost is proportional to $\frac{(\gamma-\eta)\lambda+\varepsilon}{\gamma^2-\eta^2}$. Optimizing with respect to $\gamma$ and $\eta$ shows that the balance is obtained when $\varepsilon^3=\lambda^4$ and in this case $\gamma$ is of order $\varepsilon^{1/4}$ and $\gamma-\eta$ is of order $\varepsilon^{3/4}$. The balance between the transaction cost parameters $\lambda$ and $\varepsilon$ is the only one such that both costs have the same effect on the utility function $f$.

\section{Small fixed transaction cost.}\label{FixedCost}

\subsection{Idealized no transaction cost problem}\label{NoCost}
The classical allocation problem for an investor with $z$ dollars invested in a portfolio at time $t$ is to maximize the expected utility $U$ of the portfolio dollar worth at a final time $T$; that is to find
\begin{equation}\label{Value}
	\begin{split}
	f(z,t) &= \sup E_{z,t}[U(Z(T))] \:.
	\end{split}
\end{equation}
Here the supremum is taken over all the admissible strategies available to the investor by buying or selling a risky asset. This problem was studied by Merton in \cite{merton}. We recollect the main results in order to fix the notation, which is borrowed from \cite{Goodman}. The investor's portfolio is composed of $X$ dollars in a stock whose price in dollars $S$ follows a geometric brownian motion
\begin{equation*}
	\begin{split}
	dS &= \mu S dt+\sigma S dB,
	\end{split}
\end{equation*}
and a money market account worth $Y$ dollars and growing at a constant risk free rate $r$. The SDE's for the stock and money market dollar worth are
\begin{eqnarray}
dX&=&\mu X dt +\sigma X dB \\
dY &=& r Y dt\:,
\end{eqnarray}
respectively and the SDE for the portfolio dollar worth $Z=X+Y$ is
\begin{equation}
	\begin{split}
	dZ&=[(\mu-r)X+rZ]dt+\sigma X dB\:.
	\end{split}
\end{equation}
We can write down the SDE solved by $f=f(Z,t)$ as
\begin{equation}\label{df}
	\begin{split}
	df &= \biggl\{ f_t+[(\mu-r)X+rZ] f_z +\frac{1}{2} \sigma^2 X^2 f_{zz} \biggr\}dt+\sigma X f_z dB\:.
	\end{split}
\end{equation}
The investor has the freedom to transfer money at anytime between the stock portion of his portfolio and the money market account without incurring in a transaction cost. The optimal strategy for the investor will depend only on the the time $t$ and the portfolio dollar worth $Z$; this is $X=x(Z,t)$. The Hamilton-Jacobi-Bellman equation for the optimization \eqref{Value} is
\begin{equation}
	\begin{split}
	0=\sup_{x} \biggl\{ f_t+[(\mu-r)x+rZ] f_z +\frac{1}{2} \sigma^2 x^2 f_{zz} \biggr\}\:.
	\end{split}
\end{equation} 
The solution is the Merton ratio
\begin{equation}\label{MertonRatio}
	\begin{split}
	x&=m(Z,t)=-\frac{(\mu-r)f_z}{\sigma^2 f_{zz}}\:.
	\end{split}
\end{equation}
Under the optimal control $X$ we have $f(z,t)= E_{z,t}[U(Z(T))]=E_{z,t}[f(Z(T),T)]$ and therefore the process $f(Z_t,t)$ is a martingale, which we write in differential notation as
\begin{equation}\label{martingale}
	\begin{split}
	E[df]=0\:.
	\end{split}
\end{equation}
Under a suboptimal control $X$ we will obviously have $E[df] < 0 $. The martingale property \eqref{martingale} together with equation \eqref{df} show that the value function $f$ solves the PDE
\begin{equation}\label{Merton}
	\begin{split}
	f_t-\frac{1}{2} \frac{(\mu-r)^2 f_z^2}{\sigma^2 f_{zz}}+r z f_z&=0,\quad \text{for $t<T$}\\
	f(z,T)&=U(z)\:.
	\end{split}
\end{equation}
In the rest of this paper we will denote the solution of the previous nonlinear equation by $f^0$ and make clear that it is the value function under zero transaction cost.

\subsection{Heuristic analysis of small fixed transaction cost.}\label{Heuristic}

 We will assume that the investor's broker charges a \emph{small fixed cost} of $\varepsilon$ dollars for buying or selling any number of stock shares at a fixed time. This model is only accurate for very liquid stocks and small changes in the stock portfolio. In this case the flat fee is the main source of transaction cost. Borrowing notation from \cite{Goodman}, let $L(t)$ be the dollar amount of cash spent buying stock up to time $t$, and $M(t)$ be the dollar worth of all stock sold up to time $t$. These are the controls available to the investor. The SDE's for the stock and cash components of the investor portfolio become
\begin{eqnarray}
dX&=&\mu X dt +\sigma X dB + dL-\varepsilon \delta_{dL>0} dt - dM\label{dx}\\
dY &=& r Y dt +dM-\varepsilon \delta_{dM>0} dt- dL\label{dy}\:.
\end{eqnarray}
Here $\delta_{dL>0}$ is the sum of delta functions at the times where $dL>0$ and an analogous definition holds for $\delta_{dM>0}$. At this point it is standard to perform the change of variables $\xi = X-m(Z,t)$ and work with $\xi$ instead of $X$. Here $m(Z,t)$ is the Merton ratio defined by equation \eqref{MertonRatio}. This is natural because the optimal $X$ in the case with no transaction cost is $X=m(Z,t)$. Using Ito's formula, the SDE's for $Z=X+Y$ and $\xi=X-m$ are
\begin{eqnarray}\label{dZ}
dZ&=&[(\mu-r)(m+\xi)+rZ]dt+\sigma(\xi+m)dB-\varepsilon(\delta_{dL>0} dt+\delta_{dM>0} dt)
\end{eqnarray}
and
\begin{equation}\label{dxi}
	\begin{split}
	d\xi&=[\mu(\xi+m)-m_t-\frac{1}{2} m_{zz}\sigma^2(\xi+m)^2-m_z((\mu-r)(m+\xi)+rZ)]dt\\
	&+( 1-m_z )\sigma(\xi+m) dB + \Delta \xi\:(\delta_{dM>0}+\delta_{dM>0}) dt
	\end{split}
\end{equation}
The symbol $\Delta \xi$ denotes the jump in the value of $\xi$ at time $t$. The allocation problem becomes
\begin{equation*}
	\begin{split}
	f(z,\xi,t)&= \sup E_{z,\xi,t}[U(Z(T))]\:.
	\end{split}
\end{equation*}

Our task is to perform an asymptotic analysis for the value function $f$ around the zero transaction cost value function $f^{0}$ defined in subsection \ref{NoCost}. In what follows we will give heuristic arguments that will be justified later in subsection \ref{Expansion}. We start by considering strategies such that there is a \emph{small} region of the form $\xi\in(-\gamma(Z,t),\gamma(Z,t))$ where no transaction takes place and when $\xi$ touches the boundary $\mp \gamma$ we perform a transaction in order to make $\xi=0$. This makes sense because the transaction cost is independent of the transaction amount, so the optimal strategy must make $\xi\simeq0$ immediately after  $\xi=\pm\gamma$.

The study of the dynamics of $\xi$ will be the main tool in the asymptotic analysis. The process $\xi$ has behaves like a diffusion before hitting the boundary $\pm\gamma$ and it is restricted to a confined region. All the possible values of $\xi$ are visited in a small time interval and therefore $\xi$ must reach statistical equilibrium much faster than $Z$. In other words, in all subsequent calculations we will think of $Z$ and $t$ as constants and $\xi$ will be assumed to be in equilibrium with density $u(\xi)$. The leading order of the SDE \eqref{dxi} satisfied by $\xi$ is calculated by first dropping the terms that contain an $\varepsilon$ factor. We can also drop terms that have a $\xi$ factor because $|\xi|\leq\gamma$ and we are assuming that $\gamma$ is small.  Finally, we drop the drift terms because the process stays inside a small region. Thus, up to leading order,
\begin{equation}
	\begin{split}
	d\xi\simeq adB+\Delta\xi\:,
	\end{split}
\end{equation}
with $a=\sigma (1-m_z)(\xi+m)$. 
Under such an strategy, the differential equation for the equilibrium probability density $u(\xi)$ is
\begin{eqnarray}
\frac{1}{2} a^2 u_{\xi\xi}+\frac{a^2}{\gamma^2}\delta_0&=&0
\end{eqnarray} 
with Newmann boundary conditions
\begin{eqnarray}
u_{\xi}&=&\mp \frac{1}{\gamma^2} \:\:\:\text{ at } \xi=\pm\gamma\:.
\end{eqnarray}
The constant $a^2/\gamma^2$ is the rate of particles hitting the boundary $\xi=\pm\gamma$ per unit time. When they hit the boundary, they are transported to $\xi=0$ and hence we obtain a source term proportional to $\frac{a^2}{\gamma^2}$ at $\xi=0$. The solution is
\begin{eqnarray*}
u(\xi)&=&\:\biggl(\frac{1}{\gamma}-\biggl|\frac{\xi}{\gamma^2}\biggr|\biggr)\:.
\end{eqnarray*}

The optimal boundary $\gamma$ is obtained by maximizing $E[df(Z,\xi,t)]$ over $\gamma$. Instead, we will maximize $E[df^{0}(Z,t)]$. This approximation will be justified later in subsection \ref{Expansion} . Ito's formula and equations \eqref{dZ} and \eqref{Merton} show that the idealized Merton value function process  $f^0(Z,t)$ solves the SDE
\begin{eqnarray*}
df^0&=&\biggl[ (\mu-r)\xi f^0_z+\sigma^2 m \xi f^0_{zz} +\frac{\sigma^2 \xi^2}{2} f^0_{zz}+\Delta f^0(\delta_{dL>0} +\delta_{dM>0} )\biggr] dt\\
&+&\sigma (m+\xi)f^0_z dB
\end{eqnarray*}
We will calculate the $\xi$- equilibrium expected value of the previous equation. Use that in equilibrium
\begin{eqnarray*}
E[\xi]&=&0\\
\end{eqnarray*}
and 
\begin{eqnarray*}
E[\xi^2]&=&\int_{-\gamma}^{\gamma}\xi^2u(\xi)d\xi\\
&=&\frac{1}{6}\gamma^2
\end{eqnarray*}
to get
\begin{eqnarray*}
E[df^0]&\simeq&\biggl[\frac{\sigma^2 \gamma^2}{12}a^2 f^0_{zz}+\Delta f^0 (E[\delta_{dL>0}] +E[\delta_{dM>0}] )\biggr] dt\:.
\end{eqnarray*}
\bigskip
When $dL>0$ or $dM>0$ the process $Z$ jumps from $Z$ to $Z-\varepsilon$ and therefore $\Delta f^{0}\simeq -\varepsilon f_z^{0}$. Thus,
\begin{eqnarray}\label{Expected_df}
E[df^0]&\simeq&\biggl[\frac{\sigma^2 \gamma^2}{12} f^0_{zz}-\varepsilon f_z^0 (E[\delta_{dL>0}] +E[\delta_{dM>0}] )\biggr] dt\:.
\end{eqnarray}
In order to calculate $E[\delta_{dL>0}]$, apply Ito's formula to the process $\xi^2$ to obtain the equation
\begin{eqnarray*}
d(\xi^2)\simeq2\xi a dB+a^2 dt + \Delta(\xi^2) \delta_{dL>0}dt + \Delta(\xi^2) \delta_{dM>0}dt \:.
\end{eqnarray*}
The expected value of the left hand side of the previous equation is approximately zero because $\xi$ is assumed to be close to statistical equilibrium, thus
\begin{equation*}
	\begin{split}
	0\simeq a^2 dt-\gamma^2 E[\delta_{dL>0}] dt-\gamma^2 E[\delta_{dM>0}]dt\:.
	\end{split}
\end{equation*}
The symmetry of the boundary of $(-\gamma,\gamma)$ gives
\begin{eqnarray*}
E[\delta_{dL>0}]&\simeq&\frac{1}{2\gamma^2}a^2\:,
\end{eqnarray*}
and plugin in the last equation in \eqref{Expected_df} we obtain an approximation for $\frac{E[df^0]}{dt}$ in terms of $\varepsilon$ and $\gamma$:
\begin{eqnarray*}
\frac{E[df^0]}{dt}&\simeq&\frac{\sigma^2 \gamma^2}{12} f^0_{zz}-\varepsilon f^0_z \frac{a^2}{\gamma^2} \:.
\end{eqnarray*}
Maximizing the right hand side of the previous equation over $\gamma$ gives
\begin{equation*}
	\begin{split}
	\gamma &=\biggl( -12 \frac{a^2 f_z^{0}}{\sigma^2 f_{zz}^{0}}\biggr)^{1/4} \varepsilon^{1/4}\:.
	\end{split}
\end{equation*}
and therefore
\begin{eqnarray}
\frac{E[df^0]}{dt}\simeq \biggl( - \frac{a^2\sigma^2 f_z^{0} f_{zz}^{0}}{12}\biggr)^{1/2}\varepsilon^{1/2}\:.
\end{eqnarray}
We conclude that the fixed cost $\varepsilon$ shifts the value function by $\varepsilon^{1/2}$\:.

\subsection{Fixed cost asymptotic analysis.}\label{Expansion}
The heuristic arguments of the previous section suggest that $\xi$ moves in a region of size $\varepsilon^{1/4}$. We define a rescaled variable $\widetilde{\xi}=\varepsilon^{-1/4}\xi$ but we drop the tilde to avoid cumbersome notation. We obtain from equation \eqref{dxi} that the rescaled process $\xi$ solves the SDE 
\begin{equation}\label{dxiRescaled}
	\begin{split}
	d\xi=&\varepsilon^{-1/4}[\mu(\varepsilon^{1/4}\xi+m)-m_t-\frac{1}{2} m_{zz}\sigma^2(\varepsilon^{1/4}\xi+m)^2-m_z((\mu-r)(m+\varepsilon^{1/4}\xi)+rZ)]dt\\
	+&\varepsilon^{-1/4}( 1-m_z )\sigma(\varepsilon^{-1/4}\xi+m) dB\\
	+&\Delta \xi\:.
	\end{split}
\end{equation}
Our control consists of a continuation region of the form $(-\beta,\gamma)$ and two points $\eta,\theta$. When the process $\xi$ hits the boundary $\xi=\gamma$ the investor buys stock in order to get $\xi=\eta$. Similarly, when the process $\xi$ hits the boundary $\xi=-\beta$ the investor sells stock in order to get $\xi=-\theta$. In the continuation region we have $dL=dM=0$. Using this fact, the SDE's \eqref{dxiRescaled} and \eqref{dZ} and Ito's lemma we conclude that the process $f$ solves Bellman's equation (in the continuation region):
\begin{equation}\label{fBellman}
	\begin{split}
	0=&\varepsilon^{-1/4}[\mu(\varepsilon^{1/4}\xi+m)-m_t-\frac{1}{2} m_{zz}\sigma^2(\varepsilon^{1/4}\xi+m)^2-m_z((\mu-r)(m+\varepsilon^{1/4}\xi)+rZ)]f_{\xi}\\
+&\frac{1}{2}\varepsilon^{-1/2} (1-m_z)^2\sigma^2 (\varepsilon^{1/4}\xi+m)^2 f_{\xi\xi}\\
+& \varepsilon^{-1/4}(1-m_z)\sigma^2(\varepsilon^{1/4}\xi+m)^2 f_{\xi z}\\
+&[(\mu-r)(m+\varepsilon^{1/4}\xi)+rz] f_{z} + \frac{1}{2}\sigma^2(\varepsilon^{1/4} \xi+m)^2 f_{zz}\\
+&f_t\:.
	\end{split}
\end{equation}
On the boundary of the hold region the jump terms in equations \eqref{dxiRescaled} and \eqref{dZ} dominate the Hamilton-Jacobi-Bellman equation so we must have
\begin{equation}\label{Bndry}
	\begin{split}
	f(\xi,z,t)=& f ( \eta,z-\varepsilon,t), \qquad \xi=\gamma\\
	f(\xi,z,t)=& f ( -\theta,z-\varepsilon,t), \qquad \xi=-\beta\\
	\end{split}
\end{equation}
The optimality boundary conditions (smooth pasting, take $\partial_\gamma$ and $\partial_{\eta}$ of the conditions at the boundary) are
\begin{equation}\label{OptimalBndry}
	\begin{split}
	f_{\xi}(\xi,z,t)=&0,\qquad \xi=\gamma,\eta\\
	f_{\xi}(\xi,z,t)=&0,\qquad \xi=\beta,\theta\:.
	\end{split}
\end{equation}

We propose an asymptotic expansion in powers of $\varepsilon^{1/4}$ because this is the power of $\varepsilon$ that we used to rescale $\xi$. The asymptotic expansion reads
\begin{equation}\label{expansion}
f=f^{0} (z,t)+\varepsilon^{1/4}f^{1}(\xi,z,t)+\varepsilon^{1/2}f^{2}(\xi,z,t)+\varepsilon^{3/4}f^{3}(\xi,z,t)+\varepsilon f^{4}(\xi,z,t)+o(\varepsilon)
\end{equation}
The function $f^{0}$ is the value function in the $\varepsilon=0$ case. Using the boundary condition \eqref{Bndry} and performing a Taylor expansion around $(\xi,z)$ we obtain
\begin{equation*}
	\begin{split}
	0& = -\varepsilon f_z + O(\varepsilon^2)+ f_{\xi} (\eta-\xi)+\cdots \:.
	\end{split}
\end{equation*}
This means that $f_z$ is $O(\varepsilon)$ smaller than $f_\xi$ and therefore $f^1,f^2$ and $f^3$ do not depend on $\xi$. Now we use the expansion \eqref{expansion} and collect $\varepsilon$ terms in the boundary conditions \eqref{Bndry} and obtain
\begin{equation}\label{firstLeading}
	\begin{split}
	f^{4}(\gamma,z,t)&=-f_z^{0}+f^{4}(\eta,z,t)\\
	f^{4}(-\beta,z,t)&=-f_z^{0}+f^{4}(-\theta,z,t)\\
	\end{split}
\end{equation}
Collecting $\varepsilon$ terms in the optimality boundary conditions gives
\begin{equation}\label{secondLeading}
	\begin{split}
	f^{4}_{\xi}(\xi,z,t) &= 0 ,\qquad \xi=\gamma,-\beta, \eta, -\theta\:.
	\end{split}
\end{equation}
Finally, we plug in the asymptotic expansion in Bellman's equation \eqref{fBellman}. The $O(\varepsilon^{0})$ equation involves only $f^{0}$ and it is just the Merton equation\begin{equation*}
	\begin{split}
	0 = f^{0}_t+ [(\mu-r) m+r z] f^{0}_z + \frac{1}{2}\sigma^2 m^2 f^{0}_{zz}\:. 
	\end{split}
\end{equation*}
The $O(\varepsilon^{1/4})$ equation is
\begin{equation*}
	\begin{split}
	0 &= [(\mu-r) \xi] f^{0}_z + \sigma^2 \xi m f^{0}_{zz}\\
	&+ f^{1}_t + [(\mu-r) m+r z] f^{1}_z + \sigma^2 m^2 f^{1}_{zz}\:.
	\end{split}
\end{equation*}
Since $m$ is equal to the Merton ratio the terms involving $f^{0}$ in the previous equation cancel each other and we conclude that $f^{1}$ solves Merton's equation. The final condition is $f^{1}(z,T)=0$ because $f^{0}(z,T)=f(z,\xi,T)=U(z)$ and therefore, by uniqueness,  $f^{1}(z,t)=0$. The $O(\varepsilon^{1/2})$ equation is
\begin{equation}\label{f4}
	\begin{split}
	0&=f^{2}_t+[(\mu-r)m+rz]f^{2}_z+\frac{1}{2}\sigma^2 m^2 f^{2}_{zz}+\frac{1}{2}\sigma^2 \xi^2 f^{0}_{zz}+\frac{1}{2} (1-m_z)^2\sigma^2 m^2 f^{4}_{\xi\xi}\\
	0&= K + B\xi^2+\frac{1}{2} a^2 f_{\xi\xi}^4\:,
	\end{split}
\end{equation}
where
\begin{equation*}
	\begin{split}
	K(z,t) &=f^{2}_t+[(\mu-r)m+rz]f^{2}_z+\frac{1}{2}\sigma^2 m^2 f^{2}_{zz}\\
	B(z,t) &=\frac{1}{2}\sigma^2 f^{0}_{zz}\\
	a(z,t) &=(1-m_z)\sigma m\:.
	\end{split}
\end{equation*}
Integrating with respect to $\xi$ twice we obtain
\begin{equation*}
	\begin{split}
	0&= D+C\xi+\frac{K}{2} \xi^2 + \frac{B}{12} \xi^4+\frac{1}{2} a^2 f^4\:,
	\end{split}
\end{equation*}
where $D(z,t)$ and $C(z,t)$ are constants of integration. Now we have the six equations given by  \eqref{firstLeading} and \eqref{secondLeading} and the six unknowns $\gamma,\beta,\eta,\theta,C$ and $D$. The unknown $D$ does not appear in the equations and can be set to any value that we like. Modulo $D$, the unique solution is $C=\eta=\theta=0$ and
\begin{equation*}
	\begin{split}
	\beta&=\gamma=\biggl( -12 \frac{a^2 f_z^0}{\sigma^2 f_{zz}^0}\biggr)^{1/4}\:.
	\end{split}
\end{equation*}

\section{Fixed and proportional transaction cost}\label{Both}
Assume that the fixed transaction cost is \$$\varepsilon$ for buying or selling any number of stocks  and the proportional cost is $\lambda$\% of the transaction in dollars. We will consider strategies such that $\xi=\pm \eta $ immediately after $\xi=\pm \gamma$. The new SDE's solved by $X$ and $Y$ are a small modification of equations \eqref{dx} and \eqref{dy}, namely
\begin{eqnarray}
	dX&=&\mu X dt +\sigma X dB + (1-\lambda)dL-\varepsilon \delta_{dL>0} dt - dM\label{dxBoth}\\
	dY &=& r Y dt +(1-\lambda)dM-\varepsilon \delta_{dM>0} dt- dL\label{dyBoth}\:.
\end{eqnarray}
Using Ito's formula, the SDE's for $Z=X+Y$ and $\xi=X-m$ are
\begin{equation}\label{dZBoth}
dZ=[(\mu-r)(m+\xi)+rZ]dt+\sigma(\xi+m)dB-\lambda(dL+dM)-\varepsilon(\delta_{dL>0} dt+\delta_{dM>0} dt)
\end{equation}
and
\begin{equation}\label{dxiBoth}
	\begin{split}
	d\xi&=[\mu(\xi+m)-m_t-\frac{1}{2} m_{zz}\sigma^2(\xi+m)^2-m_z((\mu-r)(m+\xi)+rZ)]dt\\
&+( 1-m_z )\sigma(\xi+m) dB + \Delta \xi
	\end{split}
\end{equation}
Applying the same heuristics used in section \ref{FixedCost} we will think of $Z$ and $t$ as constants and $\xi$ will be assumed to be in equilibrium with density $u(\xi)$. The leading order of the SDE \eqref{dxi} satisfied by $\xi$ is calculated by first dropping the terms that contain an $\varepsilon$ or $\lambda$ factor. We can also drop terms that have a $\xi$ factor because $|\xi|\leq\gamma$ and we are assuming that $\gamma$ is small.  Finally, we drop the drift terms because the process stays inside a small region. Thus, up to leading order,
\begin{equation}
	\begin{split}
	d\xi\simeq adB\label{dxiLeading}+\Delta \xi\:,
	\end{split}
\end{equation}
with $a=\sigma (1-m_z)(\xi+m)$. 
Under such an strategy, the differential equation for the equilibrium probability density $u(\xi)$ is
\begin{eqnarray}
\frac{1}{2} a^2 u_{\xi\xi}+\frac{a^2}{2(\gamma^2-\eta^2)}\delta_\eta+\frac{a^2}{2 (\gamma^2-\eta^2)}\delta_{-\eta}&=&0
\end{eqnarray} 
with Newmann boundary conditions
\begin{eqnarray}
u_{\xi}&=&\mp \frac{1}{\gamma^2-\eta^2} \:\:\:\text{ at } \xi=\pm\gamma\:.
\end{eqnarray}
The constant $\frac{a^2}{\gamma^2-\eta^2}$ is the rate of particles hitting the boundary $\xi=\pm\gamma$ per unit time.This fact can be shown, as in section \ref{FixedCost}, by applying Ito's formula to the process $\xi^2$, namely
\begin{equation*}
	\begin{split}
	d(\xi^2)\simeq2\xi a dB+a^2 dt + \Delta(\xi^2) \delta_{dL>0}dt + \Delta(\xi^2) \delta_{dM>0}dt \:.
	\end{split}
\end{equation*}
Taking the equilibrium expected value of the previous equation and using the symmetry of the boundary we obtain $E[\delta_{dL>0}]=\frac{a^2}{2(\gamma^2-\eta^2)}$. When the particles hit the boundary, they are transported to $\xi=\pm \eta$ and hence we obtain a source term proportional to $\frac{a^2}{2(\gamma^2-\eta^2)}$ at $\xi=\pm\eta$. The solution is
\begin{eqnarray*}
u(\xi)&=&\:\frac{1}{\gamma+\eta}-\biggl(\frac{|\xi|-\eta}{\gamma^2-\eta^2}\biggr)_{+}\:.
\end{eqnarray*}

The optimal boundary $\gamma$ and location $\eta$ are obtained by maximizing $E[df(Z,\xi,t)]$ over $\gamma$ and $\eta$. Instead, we will maximize $E[df^{0}(Z,t)]$. Ito's formula and equations \eqref{dZ} and \eqref{Merton} show that the idealized Merton value function process  $f^0(Z,t)$ solves the SDE
\begin{equation}
	\begin{split}
	df^0&=\biggl[ (\mu-r)\xi f^0_z+\sigma^2 m \xi f^0_{zz} +\frac{\sigma^2 \xi^2}{2} f^0_{zz}+\Delta f^0(\delta_{dL>0} +\delta_{dM>0} )\biggr] dt\\
	&+\sigma (m+\xi)f^0_z dB
	\end{split}
\end{equation}
We will calculate the $\xi$- equilibrium expected value of the previous equation. Use that in equilibrium
\begin{equation}
E[\xi]=0\:,\\
\end{equation}
and 
\begin{equation}
E[\xi^2]=\int_{-\gamma}^{\gamma}\xi^2u(\xi)d\xi=\frac{1}{6}(\gamma^2+\eta^2)\:,
\end{equation}
to get
\begin{eqnarray*}
E[df^0]&\simeq&\biggl[\frac{\sigma^2 (\gamma^2+\eta^2)}{12}a^2 f^0_{zz}+\Delta f^0 (E[\delta_{dL>0}] +E[\delta_{dM>0}] )\biggr] dt\:.
\end{eqnarray*}
When $dL>0$ or $dM>0$ the process $Z$ jumps from $Z$ to $Z-\varepsilon-\lambda|dX|$. Since $X=\xi-m(Z,t)$ when a jump takes place we have $|dX| \simeq |d\xi|=\gamma-\eta$ and therefore $\Delta f^{0}\simeq -[\varepsilon + \lambda(\gamma-\eta)]f_z^{0}$. Using that both $E[\delta_{dL>0}]$ and $E[\delta_{dM>0}]$ are equal to $\frac{a^2}{2(\gamma^2-\eta^2)}$ we obtain the equation
\begin{equation}
	\begin{split}
	\frac{E[df^{0}]}{dt}\simeq C(\gamma^2+\eta^2)+\frac{R}{\gamma^2-\eta^2}((\gamma-\eta)\lambda+\varepsilon)\:,
	\end{split}
\end{equation}
where $C=\frac{\sigma^2 f_{zz}^{0}}{12}$ and $R=-f_{z}^{0}\frac{a^2}{\gamma^2-\eta^2}$. We use the change of variables $m=\gamma+\eta$, $n=\gamma-\eta$ and write
\begin{equation*}
	\begin{split}
	\frac{E[df^0]}{dt}\simeq F(m,n)=\frac{C}{2}(m^2+n^2)+\frac{R}{mn}(n\lambda+\varepsilon)\:.
	\end{split}
\end{equation*}
In order to find the optimal values of $m$ and $n$ we calculate
\begin{equation*}
	\begin{split}
	F_m &= Cm - \frac{R}{m^2n}(n\lambda+\varepsilon)\\
	F_n &= Cn - \frac{R}{mn^2}\varepsilon\\
	\end{split}
\end{equation*}
and
$$
D^2 F=\left( \begin{array}{cc}
C+ 2\frac{R}{m^3n}(n\lambda+\varepsilon)& \frac{R\varepsilon}{m^2n^2}\\
\frac{R\varepsilon}{m^2n^2}  & C+2 \frac{R\varepsilon}{m n^3} 
 \end{array} \right)\:.
$$
Using that the gradient of $F$ is zero at the optimal controls we obtain 
\begin{equation*}
	\begin{split}
	 C^2 n^8(n\lambda+\varepsilon)&=R^2\varepsilon^{3}
	\end{split}
\end{equation*}
Now write $n=\omega \varepsilon/\lambda$ in order to get
\begin{equation}
	\begin{split}
	C^2 \omega^8(\omega+1)&=R^2\\
	\varepsilon^3&=\lambda^4\\
	m &= \frac{R}{C \omega^3}\varepsilon^{1/4}\\
	n &= \omega^3 \varepsilon^{3/4}\:.
	\end{split}
\end{equation}
We undo the change of variables and conclude that
\begin{equation*}
	\begin{split}
	\gamma &= \frac{\frac{R}{C \omega^3}\varepsilon^{1/4}+\omega^3 \varepsilon^{3/4}}{2}\\
	\eta &=\gamma - \omega^3 \varepsilon^{3/4}\:.
	\end{split}
\end{equation*}
Notice that $\varepsilon^{3} = \lambda^{4}$ is the only balance between the two types of transaction cost that makes the shift of the value function under proportional transaction cost equal to $\varepsilon^{1/2}$, which is the shift due to the proportional transaction cost.

\bibliographystyle{plain}	
\bibliography{transaction}

\begin{thebibliography}{1}

\bibitem{Atkinson}
C.~Atkinson, S.~R. Pliska, and P.~Wilmott.
\newblock Portfolio management with transaction costs.
\newblock {\em Proc. Roy. Soc. London Ser. A}, 453(1958):551--562, 1997.

\bibitem{Goodman}
J.~Goodman and D.~N. Ostrov.
\newblock Balancing small transaction costs with loss of optimal allocation in
  dynamic stock trading strategies.
\newblock {\em SIAM J. Appl. Math.}, 70(6):1977--1998, 2010.

\bibitem{janecek_shreve}
K.~Jane{\v{c}}ek and S.~E. Shreve.
\newblock Asymptotic analysis for optimal investment and consumption with
  transaction costs.
\newblock {\em Finance Stoch.}, 8(2):181--206, 2004.

\bibitem{merton}
R.~C. Merton.
\newblock {\em Continuous Time Finance}.
\newblock Blackwell, Oxford, UK, 1992.

\bibitem{atkinson_mokkhavesa}
S.~Mokkhavesa and C.~Atkinson.
\newblock Perturbation solution of optimal portfolio theory with transaction
  costs for any utility function.
\newblock {\em IMA J. Manag. Math.}, 13(2):131--151, 2002.

\bibitem{rogers}
L.~C.~G. Rogers.
\newblock Why is the effect of proportional transaction costs
  {$O(\delta^{2/3})$}?
\newblock In {\em Mathematics of finance}, volume 351 of {\em Contemp. Math.},
  pages 303--308. Amer. Math. Soc., Providence, RI, 2004.

\bibitem{shreve_soner}
S.~E. Shreve and H.~M. Soner.
\newblock Optimal investment and consumption with transaction costs.
\newblock {\em Ann. Appl. Probab.}, 4(3):609--692, 1994.

\bibitem{soner_touzi}
H.~M. Soner and N.~Touzi.
\newblock Homogenization and asymptotics for small transaction costs.
\newblock Preprint, Feb 2012.

\end{thebibliography}
\end{document}